\documentstyle[preprint,aps]{revtex}
\begin{document}
\draft
\preprint{}
\title{Reversible Pressure-Induced Amorphization in Solid
$C_{70}$ : Raman and Photoluminescence Study}
\author{N. Chandrabhas, Ajay K. Sood\cite{byline1,byline}, D.V.S. Muthu}
\address{Department of Physics, Indian Institute of Science,
Bangalore 560 012, India.}
\author{C.S. Sundar, A. Bharathi and Y. Hariharan}
\address{Materials Science Division, Indira Gandhi Centre for
Atomic Research, \\ Kalpakkam 603 102. India.}
\author{C.N.R. Rao\cite{byline}}
\address{S.S.C.U., I.I.Sc., Bangalore 560 012, India.}
\date{\today}
\maketitle
\begin{abstract}
We have studied single crystals of $C_{70}$ by Raman
scattering and photoluminescence in the pressure range from 0 to
31.1 GPa. The Raman spectrum at 31.1 GPa shows only a broad band
similar to that of the amorphous carbon without any trace
of the Raman lines of $C_{70}$. After releasing the pressure from
31.1 GPa, the Raman and the photoluminescence spectra of the
recovered sample are that of the starting $C_{70}$ crystal.
These results indicate that the $C_{70}$ molecules are stable
upto 31.1 GPa and the amorphous carbon high pressure phase
is reversible, in sharp contrast to the results on
solid $C_{60}$. A qualitative explaination is suggested in terms
of inter- versus intra-molecular interactions.
\end{abstract}


\narrowtext

Among the many fascinating properties of fullerenes, studies
on structure of $C_{60}$ and $C_{70}$ and their structural
transformations as a function of temperature and pressure have
attracted considerable attention \cite{krish}. The orientational
order-disorder transition occurs at T$_c$ $\sim$ 250 K in $C_{60}$
\cite{heiney} and the
transition temperature increases with pressure ($\sim$ 100 K/GPa)
resulting in the room-temperature phase transition at $\sim$ 0.4 GPa
\cite{samara}. The stability of the $C_{60}$ buckyballs at high
pressures and the nature of the compressed phase have been
debated extensively in the last two years
\cite{duclos,raptis,snoke,moshary,snoke2,regueiro}.
Early x-ray diffraction experiments \cite{duclos} showed a structural phase
transition of cubic $C_{60}$ to a lower symmetry phase at $\sim$
20 GPa under non-hydrostatic compression. Raman Spectroscopy
\cite{raptis,snoke} and optical reflectivity \cite{snoke}
studies showed that the solid $C_{60}$ undergoes an irreversible
transformation to an amorphous carbon phase at pressures greater
than 22 GPa.  Moshary {\sl et al.} \cite{moshary} reported
anomalously high transparency of the irreversible high pressure
phase thus suggesting the collapse of the $C_{60}$ molecules
into a new structure of carbon. However, the recent measurements
\cite{snoke2} of optical absorption at high pressures do not find
any evidence of high transparency and reconfirm the earlier
conclusion of irreversible transformation of $C_{60}$
to amorphous carbon. It has also been reported \cite{regueiro} that rapid,
non-hydrostatic compression of solid $C_{60}$ transforms it to
cubic diamond at about 20 $\pm$ 5 GPa. Such a conversion to the
cubic diamond has not been seen in hydrostatic pressure
experiments \cite{snoke2}.

Compared to $C_{60}$, the stability of $C_{70}$ - the second
most abundant fullerene present in the arc-processed carbon
deposits, with respect to compression and the nature of the high
pressure phase is relatively much less understood. The order-disorder
transitions occur at $\sim$ 276 K and 337 K \cite{vaughan,ramasesha}.
Recent high pressure x-ray diffraction experiments \cite{christides} at room
temperature reveal fcc $\rightarrow$ rhombohedral phase
transition at 0.35 GPa with fcc phase coexisting till $\sim$ 1
GPa. On further compression beyond 18 GPa, solid $C_{70}$
transforms to an amorphous phase. Though it has been stated
\cite{christides} that the transition is irreversible, there are
no results on decompression and on  the pressure-cycled
recovered sample.

In this paper we report our high pressure
Raman and photoluminescence (PL) results on single crystals of
$C_{70}$ upto 31.1 GPa. Raman lines, characteristic of
intramolecular modes of $C_{70}$ could be followed upto 12 GPa. The
lines broaden and shift with pressure, with a characteristic change
of slope at $\sim$ 1 GPa which can be attributed to orientational ordering
transition. At P $>$ 20 GPa, the Raman spectra starts showing a broad
band near 1650 cm$^{-1}$, matching very well with the high pressure
phase of the amorphous carbon \cite{goncharov} and there are no
Raman signatures corresponding to the intramolecular vibrations
of the $C_{70}$. The most spectacular result is that the Raman spectra at 0.1
GPa
in the decreasing pressure run and in the recovered sample
(outside the diamond anvil cell(DAC)) are that of the starting
$C_{70}$ crystal. In addition to Raman studies, PL measurements have
also been carried out as a function of pressure. The PL of the
pressure-cycled sample is seen to be similar to that of the starting
$C_{70}$ sample. These results unambigously indicate that the
amorphous phase at high pressures is reversible and the
$C_{70}$ molecules are stable upto 31.1 GPa, in sharp contrast
to the high pressure results on the solid $C_{60}$.

The mixture of $C_{60}$ and $C_{70}$ was prepared by the
well-known contact-arc vaporization of graphite in helium
atmosphere.  The soxhlet extract was subjected to repeated
chromatographic separation using neutral grade alumina column
using toluene and hexane as eluent. Based on the UV-visible absorption,
infrared and Raman spectra, the purity of $C_{70}$ is estimated to be
better than 99\%. Small single crystals were grown by the
temperature gradient vapour deposition method. High pressure Raman and PL
experiments were carried out using 5145 \AA \/ argon laser line
(power $<$ 5 mW before the DAC) at room temperature in a gasketed
Mao-Bell type DAC with 4:1 methanol-ethanol mixture as
pressure-transmitting medium. Pressure was measured by the
well-known ruby fluorescence technique. The interference of the
ruby fluorescence ($\sim$ 1.78 eV to 1.67 eV at ambient pressure)
with the PL of the $C_{70}$ crystal was minimised by keeping the
ruby crystal (size $\sim$ 20 $\mu$m) as far away from the sample
as possible in the gasket hole (size $\sim$ 200 $\mu$m). The
spectra were recorded using a DILOR-XY spectrophotometer
equipped with a liquid nitrogen cooled CCD detector as well as
using a Spex Ramalog with a cooled photomultiplier tube
(Burleigh C31034A). Twelve pressure runs
were done reaching a maximum pressure of 31.1 GPa. The ruby
fluorescence lines were seen to broaden beyond 12 GPa but were
clearly resolved upto the highest pressure.

We shall first discuss the Raman results. In the spectral range of
200 - 1700 cm$^{-1}$, we observe 20 distinct intramolecular modes at almost
ambient
pressure ($<$ 0.1 GPa) with the sample in the DAC, similar to
the earlier reports \cite{van,bhas}. The region 1300 - 1400
cm$^{-1}$ is dominated by the first order Raman line of the
diamond anvils at 1332 cm$^{-1}$. The variation of Raman lines
with increase in pressure in the range 0 to 31.1 GPa are shown
in Fig. 1(a) and (b). As the pressure increases, $C_{70}$ Raman lines
broaden and their intensities decrease
substantially. Even the 1567 cm$^{-1}$ mode which is most
intense at ambient pressures can be followed only upto $\sim$ 12
GPa after which it merges with the background (see Fig. 1(a)).  We find that
the
slopes in the pressure dependence of the peak position of Raman
modes at 1567, 1516, 1470 and 1449 cm$^{-1}$ change at $\sim$ 1
GPa \cite{sood2}, which can be related to the reported orientational
ordering phase transition from fcc to rhombohedral structure
observed by x-ray diffraction \cite{christides}.

At high pressures (P $>$ 20 GPa), only a broad
Raman band appears between 1500 and 1900 cm$^{-1}$. The intensity of
this band is very weak. In order to improve signal to noise ratio,
these measured spectra (at P $>$ 20 GPa) were subjected to discrete
wavelet transform filtering \cite{daubachies}. The Raman
spectrum at pressure 31.1 GPa shows a broad band peaked at 1720 cm$^{-1}$.
This broad band is exactly similar to the corresponding high
pressure Raman spectrum of the amorphous carbon which has a Raman
band centered about $\omega = 1580$ cm$^{-1}$ at ambient pressure
and pressure derivative d$\omega$/dP of 4.4 cm$^{-1}$/GPa
\cite{goncharov}. On decrease of pressure from 31.1 GPa, the
Raman measurements at 10 GPa still show a broad band at 1600
cm$^{-1}$ (see inset Fig. 2), implying that the
amorphous phase is present upto this pressure. The most
interesting result is that the Raman spectra at 0.1 GPa and that
of the recovered sample outside the DAC are that of the starting
$C_{70}$ crystal wherein 20 lines are again seen in the range of
200 - 1700 cm$^{-1}$.

Fig.\ \ref{third}(a) shows the PL spectra of the $C_{70}$ crystal
in the DAC at four different pressures of 0, 0.7, 1.7 and 2.8
GPa in the increasing pressure run. The PL spectrum of the starting $C_{70}$
crystal
is similar to the one reported earlier of the $C_{70}$ film on
Si substrate \cite{shin}. The PL bands move to lower energies with
increasing pressure, similar to that in $C_{60}$. At P $>$ 2.8
GPa, the PL band shifts to less than 1.4 eV, the lower limit of
detection in our experiments. The red-shift of the PL band with
pressure ($\sim$ - 0.09 eV/GPa) is
mostly associated with the reduction of the band gap which, in
turn, is related to the broadening of the valence and
conduction bands due to pressure-induced enhancement of
intermolecular interactions \cite{sood}.

The PL of the sample recovered after cycling to 31.1 GPa
(Fig.\ \ref{third}(b)) is very similar to the starting sample, except for a
small blue shift of 0.1 eV. The blue shift which reflects a decrease in
overlap between $C_{70}$ orbitals can be either due to defects
or clathration of the alcohol molecules in
$C_{70}$, as seen in the case of $C_{60}$ crystals \cite{kamaras}. The results
shown in Fig.\ \ref{second} and Fig.\ \ref{third}(b) unambiguously
show that the amorphous phase as identified by Raman spectra at high
pressures beyond 20 GPa (cf Fig. 1(b)) reverts back to the solid $C_{70}$ on
release of pressure. This is in sharp contrast to the irreversible
amorphization of solid $C_{60}$ at P $>$ 22 GPa. In the case of
graphite (the parent material for fullerenes), transformation to an
amorphous phase occurs at 23 GPa \cite{goncharov2} which has been reported to
be
reversible on decompression \cite{snoke}.

At this stage, we address the question if the possible quasi-hydrostatic
nature of the pressure in the DAC can influence the reversible
nature of the high pressure phase. Broadened linewidths and the
separation ($\Delta$) of the R$_1$ and R$_2$ ruby flurescence
lines are generally regarded as indicators of the presence of non
hydrostatic stress components. In our experiments, the R$_1$ and
R$_2$ lines are clearly resolved even at the highest pressure of
31.1 GPa, although their linewidths are broadened for P $>$ 12
GPa. The pertinent observation in relation to our results is
that in the return pressure run, the linewidth and the
separation $\Delta$ for P $=$ 10 GPa and 0.1 GPa remain nearly
similar whereas the measured Raman spectra are drastically
different. This indicates that the nature of pressure, in
particular the extent of non-hydrostaticity, does not
influence our main results.

We can attempt to rationalise the pressure behaviour seen in
$C_{70}$ in comparison with those observed earlier
in $C_{60}$ \cite{snoke} and graphite \cite{goncharov2,snoke} based on
intramolecular distances and steric constraints \cite{sikka}. It
must be remarked that pressure-induced amorphization is also
seen in other molecular crystals like
SiO$_{2}$ \cite{hemley}, AlPO$_{4}$ \cite{kruger} and LiCsSO$_{4}$
\cite{shashikala}. Some of these systems like
SiO$_{2}$ retain their amorphous nature on pressure release whereas
other like AlPO$_{4}$ and LiCsSO$_{4}$ recrystallise. In this picture
pressure reduces the nearest neighbour distance between the molecules
and below a limiting distance, the intermolecular interactions can
become comparable to intramolecular interactions themselves, resulting in the
distortion of the molecular units and the formation of a new phase having
different bonding and structure.

The nearest carbon-carbon distance between the neighbouring bucky
balls is d$_{C-C}$ = d$_{n-n}$ - $\sigma$, where d$_{n-n}$ is the
nearest neighbour distance between the centres of the molecules and
$\sigma$ is the relevent molecular dimension. The $C_{60}$ molecules are
spherical with $\sigma$ = 7.06 \AA \/ \cite{heiney}. The $C_{70}$ molecules, on
the
other hand are ellipsoidal \cite{taylor} with long axis of 7.916 \AA \/ and
short-axis of 7.092 \AA \/ and are oriented in the high pressure
rhombohedral phase (P $>$ 1 GPa) with the long axis along the [111] direction
and
therefore, in the close packing (110) plane, $\sigma$ = 7.092 \AA. \/ The
pressure dependence of
d$_{C-C}$, related to d$_{n-n}$ has been obtained from the Murnaghan
equation of state \cite{murnaghan} using the known bulk modulus
and its pressure-derivative \cite{duclos,christides,hanfland}. For
$C_{60}$ solid, d$_{C-C}$ reduces from 3.04 \AA \/ (P = 0) to
1.89 \AA \/ at 22 GPa whereas for $C_{70}$ solid it decreases
from 3.39 \AA \/ (P = 0) to only about 2.67 \AA \/ at 20 GPa. In
graphite, interlayer separation along the c-axis (equivalent to
d$_{C-C}$) reduces from 3.35 \AA \/ (P = 0) to 2.76 \AA \/ at 23
GPa. The decrease of d$_{C-C}$ is steeper in $C_{60}$ than that
of $C_{70}$ and graphite. Theoretical calculations in $C_{60}$
indicate \cite{xu} that as
d$_{C-C}$ decreases, the intermolecular interaction, which is van-der
Waal type at ambient pressure, acquires some covalent character
with an associated partial conversion from sp$^{2}$ to sp$^{3}$ hybridization.
When
the fraction of sp$^{3}$ bonds increases at high pressures, the
structure can be close to amorphous carbon which is characterised by
the presence of $\sim$ 15\% sp$^{3}$ bonding \cite{galli}. A
similar mechanism could account for the amorphisation of
$C_{70}$ with pressure. The question of
reversibility of the high pressure phase on decompression should be
related to the relative strengths of the inter- versus
intramolecular interactions. If the latter is sufficiently stronger, the
molecules will not be permanently distorted/destroyed and the high
pressure amorphous phase will be reversible on decompression. This seems to be
the
case in solid $C_{70}$ where the contact intermolecular distance
(d$_{C-C}$ = 2.67 \AA \/ at 20 GPa) is much larger than in the high
pressure phase of the $C_{60}$ (d$_{C-C}$ = 1.89 \AA \/ at 22 GPa).
While we have provided a qualitative explanation as to why reversible
transition is observed in $C_{70}$
in contrast to $C_{60}$, the present results should motivate detailed
molecular dynamics simulation to understand the high pressure phase
of $C_{70}$. It will be worthwhile to extend the measurements to
higher pressures to explore the possibility of irreversible
transformations, if any, in $C_{70}$.

In conclusion, we have shown that the high pressure phase of the
solid $C_{70}$ has Raman signatures of the amorphous carbon. The
amorphous phase reverts to the crystalline $C_{70}$ on
decompression as evident by the Raman lines associated with
the intramolecular vibrations of $C_{70}$ and the
PL, implying that the $C_{70}$ molecules are
stable upto 31.1 GPa. Our experiments also
suggest a need to do careful high pressure x-ray diffraction
experiments with particular emphasis on the decreasing pressure
cycle and the pressure-cycled recovered samples. Further, high
pressure experiments on higher fullerenes will be interesting to
see if solid $C_{70}$ is unique in its ability to withstand
pressure without irreversible transformation. The
present experiments on $C_{70}$, in conjuction with those in
$C_{60}$ and graphite should help to obtain an understanding of
the nature of C-C interactions and the relative stability of
different forms of carbon.

\acknowledgments

One of us (AKS) thanks Department of Science and Technology for
the financial support.

\begin{figure}
\caption{(a) Raman spectra in the range of 1400 - 1700 cm$^{-1}$
at different pressures upto 15.3 GPa in the increasing
pressure cycle. Magnification factors are indicated. (b) Raman
spectra in the spectral range 1400 - 1900 cm$^{-1}$ at higher
pressures. The thick solid lines are the wavelet transforms of
the corresponding recorded spectra, shown by thin lines.}
\label{first}
\end{figure}

\begin{figure}
\caption{Raman spectra of $C_{70}$ cycled to 31.1 GPa during the
decreasing pressure run. The inset shows the spectra at 10 GPa
and 0.1 GPa, note that the full spectrum of the recovered sample
shows all the characteristic Raman lines of the starting
$C_{70}$ sample.}
\label{second}
\end{figure}

\begin{figure}
\caption{Photoluminescence spectra (a) at four typical
presssures in the increasing pressure cycle and (b) of the recovered
sample after cycling to 31.1 GPa, simultaneously recorded with
Raman spectrum shown in Fig. 2 (P = 0(Recovered)). Asteriks in top
curve of (a) mark the contributions from the ruby fluorescence.}
\label{third}
\end{figure}

\begin{references}
\bibitem[*]{byline1} Author to whom correspondence should be addressed.
\bibitem[\ddag]{byline}  Also at Jawaharlal Nehru Centre for Advanced
Scientific Research, IISc. Campus, Bangalore 560 012, India.
\bibitem{krish} For a review, see H.R. Krishnamurthy and A.K.
Sood, Rev. Solid State Sci. {\bf 5}, 587(1991); J.E. Fisher and P.A.
Heiney, J. Phys. Chem. Solids {\bf 54}, 1725(1993).
\bibitem{heiney} P.A. Heiney, J.E. Fisher, A.R. McGhie, W.J.
Romanow, A.W. Denenstein, J.P. McCauley Jr., A.B. Smith III and
D.E. Cox, Phys. Rev. Letters {\bf 66}, 2911(1991).
\bibitem{samara} G.A. Samara, J.E. Schirber, B. Morosin,
L.V.Hansen, D. Loy and A.P. Sylwester, Phys. Rev. Letters {\bf
67}, 3136(1991).
\bibitem{duclos} S.J. Duclos, K. Brister, R.C. Haddon, A.R.
Kortan and F.A. Thiel, Nature {\bf 351}, 380(1991).
\bibitem{raptis} Y.S. Raptis, D.W. Snoke, K. Syassen, S. Roth,
P.Bernier and A. Zahab, High Pressure Research {\bf 9}, 41(1992).
\bibitem{snoke} D.W. Snoke, Y.S. Raptis and K. Syassen, Phys.
Rev. {\bf B45}, 14419(1992).
\bibitem{moshary} F. Moshary, N.H. Chen, I.F. Silvera, C.A.
Brown, H.C. Dorn, M.S. de Vries and D.S. Bethune, Phys. Rev.
Letters, {\bf 69}, 466(1992).
\bibitem{snoke2} D.W. Snoke, K. Syassen and A. Mittelbach, Phys.
Rev. {\bf B47}, 4146(1993).
\bibitem{regueiro} M.N. Regueiro, P. Monceau and J. Hodeau,
Nature {\bf 355}, 237(1992).
\bibitem{vaughan} G.B.M. Vaughan, P.A. Heiney, J.E. Fischer,
D.E. Luzzi, D.A. Ricketts-Foot, A.R. McGhie, W.J. Romanow, B.H.
Allen, N. Coustel, J.P. McCauley Jr., and A.B. Smith III,
Science {\bf 254}, 1350(1991).
\bibitem{ramasesha} Another transition at 325 K is observed in
S.K. Ramasesha, A.K. Singh, R. Sheshadri, A.K. Sood and C.N.R.
Rao, Chem. Phys. Letters {\bf 220}, 203(1994).
\bibitem{christides} C. Christides, I.M. Thomas, T.J.S. Dennis
and K. Prassides, Europhys. Letters {\bf 22}, 611(1993).
\bibitem{goncharov} A.F. Goncharov and V.D. Andreev, Sov. Phys.
JETP {\bf 73}, 140(1991).
\bibitem{van} P.H.M. van Loosdrecht, M.A. Verheijen, H. Meekes,
P.J.M. van Benthum and G. Meijer, Phys. Rev. {\bf B47},
7610(1993) and references therein.
\bibitem{bhas} N. Chandrabhas, K. Jayaram, D.V.S. Muthu, A.K.
Sood, R. Seshadri and C.N.R. Rao, Phys. Rev. {\bf B47}, 10963(1993).
\bibitem{sood2} A.K. Sood, N. Chandrabhas, D.V.S. Muthu, Y.
Hariharan, A. Bharathi and C.S. Sundar, Phil. Mag. (to be published).
\bibitem{daubachies} J. Daubachies, Wavelets (SIAM, Philadelphia)
1992; M.C. Valsakumar (unpublished).
\bibitem{shin} E. Shin, J. Park, M. Lee, D. Kim, Y.D. Suh, S.I.
Yang, S.M. Jin and S.K. Kim, Chem. Phys. Letters {\bf 209},
427(1993).
\bibitem{sood} A.K. Sood, N. Chandrabhas, D.V.S. Muthu, A.
Jayaraman, N. Kumar, H.R. Krishnamurthy, T. Pradeep and C.N.R.
Rao, Solid State Communications, {\bf 81}, 89(1992).
\bibitem{kamaras} K. Kamaras, A. Breitschwerdt, S. Pekker, K.
Fodor-Csorba, G. Faigel and M. Tegze, Appl. Phys. {\bf A 56},
231(1993).
\bibitem{goncharov2} A.F. Goncharov, JETP Letters, {\bf 51},
418(1990).
\bibitem{sikka} S.K. Sikka and S.M. Sharma, Current Science {\bf 63},
317(1992); S.K. Sikka, S.M. Sharma and R. Chidambaram, in the Proc.
XIV AIRAPT Conf., Colorado Springs, USA, 1993.
\bibitem{hemley} R.J. Hemley, A.P. Jephcoat, H.K. Mao, L.C. Ming and
M.H. Manghnani, Nature {\bf 334}, 52(1988).
\bibitem{kruger} M.B. Kruger and R. Jeanloz, Science {\bf 249},
647(1990).
\bibitem{shashikala} M.N. Shashikala, N. Chandrabhas, K. Jayaram, A.
Jayaraman and A.K. Sood, J. Phys. Chem. Solids {\bf 55}, 107(1994).
\bibitem{taylor} R. Taylor, J.P. Hare, A.K. Abdul-Sada and H.W.
Kroto, J. Chem. Soc. Chem. Commun. {\bf 20}, 1423(1990).
\bibitem{murnaghan} F.D. Murnaghan, Proc. Natl. Acad. Sci. {\bf 30},
244(1947).
\bibitem{hanfland} M. Hanfland, H. Beister and K. Syassen, Phys.
Rev. {\bf B 39}, 12598(1989).
\bibitem{xu} Y-N. Xu, M-Z. Huang and W.Y. Ching, Phys. Rev. {\bf B
46}, 4241(1992).
\bibitem{galli} G. Galli, R.M. Martin, R. Car and M. Parrinello,
Phys. Rev. Letters {\bf 62}, 555(1989).
\end{references}
\end{document}